\documentclass[12pt,preprint]{emulateapj}
\journalinfo{\textsc{The Astrophysical Journal}, in press}
\submitted{Received 2005 January 31; accepted 2005 February 21}

\usepackage{natbib}
\usepackage{lscape}

\newcommand{\xmm}{\mbox{{\em XMM-Newton\/}}}
\newcommand{\rosat}{\mbox{{\em ROSAT\/}}}
\newcommand{\fxunits}{\mbox{ergs cm$^{-2}$ s$^{-1}$}}
\newcommand{\lxunits}{\mbox{ergs s$^{-1}$}}

\newcommand{\xmmu}{\mbox{XMMU\,J2235.3--2557}}

\def\down#1{\leavevmode \lower.70ex\hbox{#1}}
\def\lesssim{\mathrel{\down{$\buildrel < \over \sim$}}}
\def\gtrsim{\mathrel{\down{$\buildrel > \over \sim$}}}

\begin{document}

\title{Discovery of an \mbox{X-ray}--Luminous Galaxy Cluster at
$\MakeLowercase{z}=1.4$\altaffilmark{1,2}}

\author{C.R. Mullis\altaffilmark{3,4},
        P. Rosati\altaffilmark{4},
	G. Lamer\altaffilmark{5},
	H. B\"{o}hringer\altaffilmark{6},
	A. Schwope\altaffilmark{5},
        P. Schuecker\altaffilmark{6} and
	R. Fassbender\altaffilmark{6}}

\altaffiltext{1}{Based on observations obtained with {\em XMM-Newton}, an ESA mission with contributions from NASA}

\altaffiltext{2}{Based on observations obtained at the European
Southern Observatory using the ESO Very Large Telescope on Cerro
Paranal (ESO programs 72.A-0706, 73.A-0737, 74.A-0023 and
274.A-5024) }
\altaffiltext{3}{University of Michigan, Department of Astronomy, 
                 918 Dennison Building, Ann Arbor, MI 48109-1090, 
                 cmullis@umich.edu}
\altaffiltext{4}{European Southern Observatory, Headquarters,
                 Karl-Schwarzschild-Strasse 2, Garching bei M\"unchen
                 D-85748, Germany} 
\altaffiltext{5}{Astrophysikalisches Institut
                 Potsdam, An der Sternwarte 16, 14482 Potsdam,
                 Germany}
\altaffiltext{6}{Max-Planck
                 Institut f\"u{r} extraterrestrische Physik,
                 Giessenbachstrasse 1603, Garching, D-85741, Germany}  
\shorttitle{X-RAY--LUMINOUS GALAXY CLUSTER AT $z=1.4$}
\shortauthors{MULLIS ET AL.}

\begin{abstract}

We report the discovery of a massive, \mbox{X-ray}-luminous cluster of
galaxies at \mbox{$z$=1.393}, the most distant \mbox{X-ray}--selected
cluster found to date.  \xmmu~was serendipitously detected as an
extended \mbox{X-ray} source in an archival \xmm~observation of
NGC\,7314.  VLT-FORS2 $R$ and $z$ band snapshot imaging reveals an
over-density of red galaxies in both angular and color spaces.  The
galaxy enhancement is coincident in the sky with the \mbox{X-ray}
emission; the cluster red sequence at \mbox{$R-z \simeq 2.1$}
identifies it as a high-redshift candidate.  Subsequent VLT-FORS2
multi-object spectroscopy unambiguously confirms the presence of a
massive cluster based on 12 concordant redshifts in the interval
\mbox{$1.38 < z < 1.40$}.  The preliminary cluster velocity dispersion
is \mbox{$762 \pm 265$ km~s$^{-1}$}.  VLT-ISAAC $Ks$ and $J$ band
images underscore the rich distribution of red galaxies associated
with the cluster.  Based on a 45\,ks \xmm~observation, we find the
cluster has an aperture-corrected, unabsorbed \mbox{X-ray} flux of
$f_{\rm X}=(3.6 \pm 0.3) \times 10^{-14}$ \fxunits, a rest-frame
\mbox{X-ray} luminosity of $L_{\rm X}=(3.0 \pm 0.2) \times
10^{44}~h_{70}^{-2}$ \lxunits~(0.5--2.0 keV), and a temperature of
$kT=6.0^{+2.5}_{-1.8}$\,keV. Though \xmmu~is likely the first
confirmed $z > 1$ cluster found with \xmm, the relative ease and
efficiency of discovery demonstrates that it should be possible to
build large samples of $z > 1$ clusters through the joint use of X-ray
and large, ground-based telescopes.

\end{abstract}

\keywords{galaxies: clusters: general --- \mbox{X-ray}s: general}

\section{Introduction} 
\label{Introduction} 

There is a strong impetus in astronomy to discover and investigate
objects at ever increasing redshifts in order to probe the state of
the Universe at increasingly earlier stages of cosmic history.  Such
observations allow us to construct evolutionary sequences which
ultimately reveal the underlying mechanisms and parameters that define
the Universe.  The high-redshift push is acutely applicable to the
study of galaxy clusters.  Their density evolution and distribution on
large scales are very sensitive to the cosmological
framework. Furthermore, clusters play a key role in tracking the
formation and evolution of massive early-type galaxies. It is
important to recognize that the leverage on both the derived
cosmological parameters and the efficacy of evolutionary studies is
greatly enhanced as we probe to higher redshifts.

\mbox{X-ray} selection is currently the optimal technique for
constructing large well-defined samples of distant clusters
\citep*[see review by][]{Rosati2002b}. However infrared
large-area surveys may well become a complementary approach
\citep[e.g.,][]{Eisenhardt2004}.  The present status of X-ray cluster
samples is due in large part to numerous \rosat-based surveys.
We now have definitive local samples ($z \lesssim 0.3$) totaling
$\sim$1000 clusters \citep[e.g., \mbox{REFLEX};][]{Boehringer2004} and
high-redshift samples totaling a few hundred clusters \citep[e.g.,
160SD;][]{Vikhlinin1998,Mullis2003}.  Galaxy clusters were routinely
discovered at $z>0.5$, and occasionally at $z>0.8$.  However, the
$z>1$ domain has been largely unexplored.  Only five X-ray--emitting
clusters are known here
\citep{Stanford1997,Rosati1999,Stanford2002,Rosati2004,Hashimoto2004};
four of which are from the RDCS survey of \citet{Rosati1998}.

\begin{figure*}
\epsscale{1.2}
\plotone{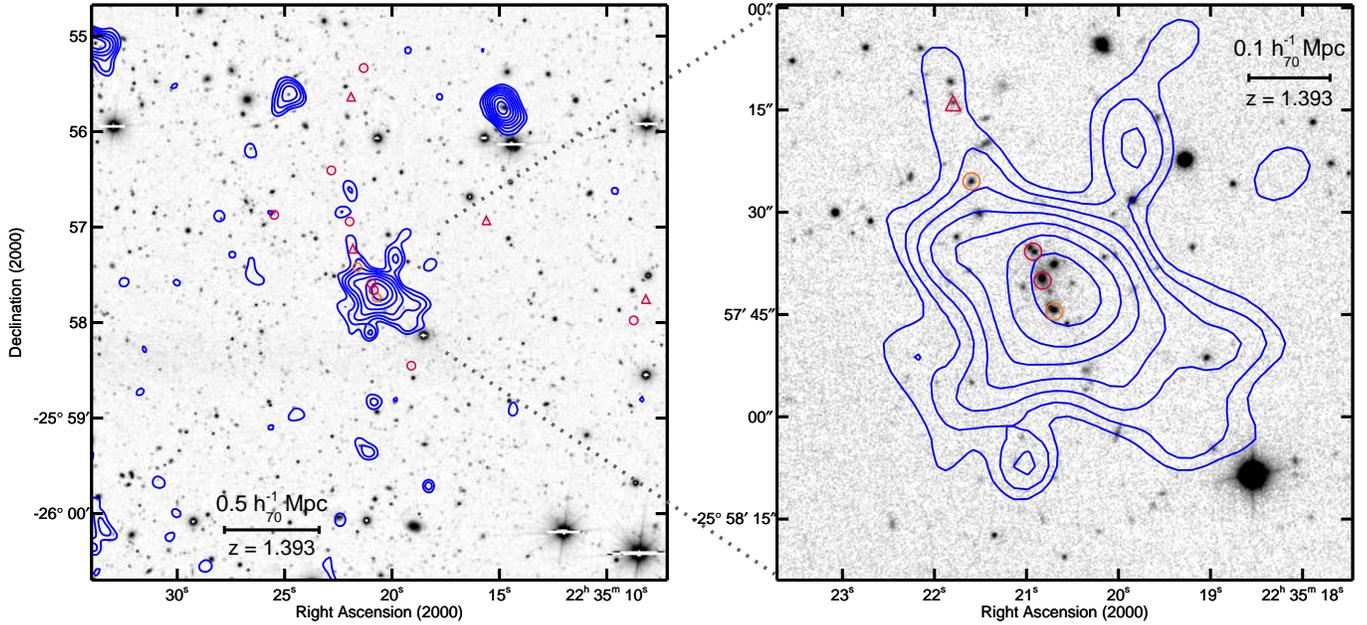}
\caption{Galaxy cluster \xmmu~at $z=1.393$ --- {\em Left:} VLT-FORS2
$R$-band image (1140s) overlaid with X-ray contours from a 45\,ks
\xmm~observation.  The \mbox{0.5--2.0\,keV} X-ray image from the EPIC
M1+M2 detectors has been smoothed with a 4\arcsec~Gaussian kernel;
eight logarithmically-spaced contours are drawn between 0.2 and 1
count per 2\arcsec~pixel. The prominent \mbox{X-ray} point source
north-west and 2.3\arcmin~further off-axis than the cluster is a
Seyfert 2 galaxy at $z=0.4060$. {\em Right:} VLT-ISAAC $Ks$ image
(3600s) overlaid with the same \mbox{X-ray}
contours. Spectroscopically confirmed members \mbox{($1.38 < z <
1.40$)} are marked in red.  Two galaxies at \mbox{$1.37 < z < 1.38$}
are marked in orange.  In both cases circles indicate absorption line
galaxies and triangles indicate emission line galaxies.\label{fig:images}}
\end{figure*}

It is now possible to redress the lack of knowledge of galaxy clusters
at $z > 1$ using \xmm, which features unprecedented sensitivity, high
angular resolution and wide-field coverage.  Several general surveys
are underway \citep[e.g.,][]{Romer2001,Pierre2004,Schwope2004}.  We
briefly describe here the first high-redshift discovery resulting from
our pilot program, which is specifically focused on the identification
of $z > 1$ galaxy clusters using \xmm.

\begin{figure*}
\epsscale{0.89}
\epsscale{.925}

\plotone{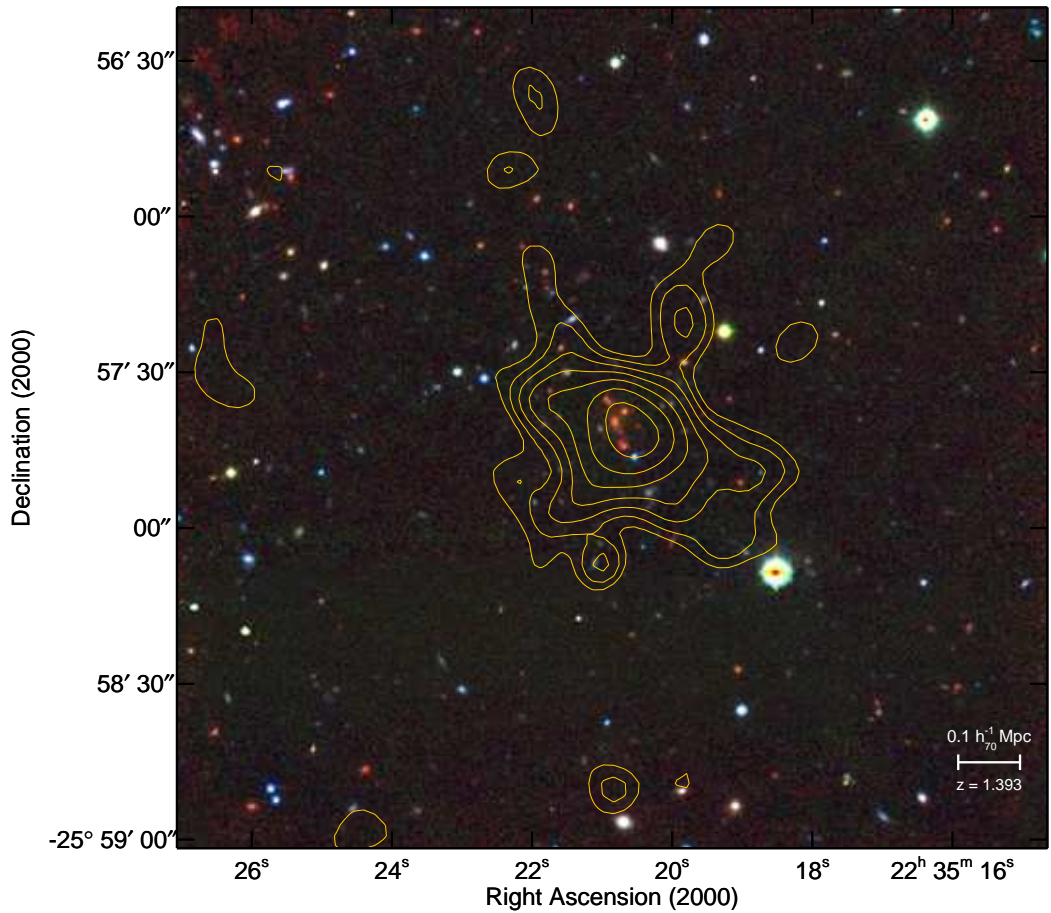}

\caption{Color image of \xmmu~overlaid with the same X-ray contours as
\mbox{Figure \ref{fig:images}}. The red channel is a VLT-ISAAC $Ks$
image (3600s); the green channel is a VLT-FORS2 $z$-band image (480s);
the blue channel is a VLT-FORS2 $R$-band image
(1140s).\label{fig:rzk}}
\end{figure*}


\section{\mbox{X-ray} Selection and Analysis}
\label{x-ray}

We have initiated a search for distant, X-ray luminous clusters
through the serendipitous detection of extended X-ray emission in
archival \xmm~observations with exposure times $>$20\,ks.  Our ultimate
objective is to construct an X-ray flux-limited sample of tens of
galaxy clusters at \mbox{$z \gtrsim 1$}.  A more immediate goal has
been to develop a rapid and efficient observational strategy to
identify the most distant systems ($z > 1 $).  One of the noteworthy
objects identified in our initial test fields is
\mbox{XMMU\,J2235.3-2557} which is detected in a 45\,ks
\xmm~observation of the Seyfert 1.9 galaxy NGC~7314.  The source is
located at 7.7\arcmin~off-axis in the observation recorded on 2 May
2001 (obsid 0111790101). As demonstrated by the \mbox{X-ray} flux
contours in \mbox{Figure \ref{fig:images}}, this source is extended on
arcminute scales and is clearly resolved in comparison to the
prominent X-ray point source to the north-west and 2.3\arcmin~further
off-axis.  The \mbox{X-ray} centroid of \xmmu~in equatorial
coordinates is \mbox{$\alpha_{\rm J2000.0}=22^{\rm h}35^{\rm
m}20.6^{\rm s}, \delta_{\rm J2000.0}=-25^{\circ}57^{\rm m}42^{\rm s}$}
which corresponds to a Galactic latitude of $b=-59.6^{\circ}$.
Extended X-ray sources at high Galactic latitudes are almost
exclusively galaxy clusters.

Our X-ray analysis is restricted to the two EPIC-MOS detectors since
the EPIC-pn detector was operated in small window mode
to avoid pile-up and to permit rapid variability observations of
NGC\,7314, and thus did not image the location of \xmmu.  The
available data are equivalent to a $\sim$22.5\,ks observation with all
three detectors.  An effective integration time of 38\,ks remains
after screening periods of high background.  Counts were extracted
from a 50\arcsec~radius circular region centered on the source; the
background was estimated locally using three source-free circular
apertures ($r=60$\arcsec$-$120$\arcsec$) flanking \xmmu.  There are
280 net source counts in the \mbox{0.3--4.5\,keV} band for the
combined MOS detectors (M1+M2).  This corresponds to an unabsorbed
aperture flux of \mbox{($2.6 \pm 0.2) \times 10^{-14}$ ergs cm$^{-2}$
s$^{-1}$} in the \mbox{0.5--2.0\,keV} energy band modeling
the source with a 6\,keV thermal spectrum (details presented in
\S\,\ref{discussion}).  Measurement errors are given at the 68\%
confidence interval ($1\sigma$) throughout.

\xmmu~was also serendipitously detected in a \rosat~PSPC observation
\citep[1WGA\,J2235.3-2557;][]{White1994}.  The \rosat~flux of
\mbox{($2.4 \pm 0.4) \times 10^{-14}$ ergs cm$^{-2}$ s$^{-1}$}
(\mbox{0.5--2.0\,keV}) is in excellent agreement with our \xmm~result.
This source was not followed up by the \rosat-era cluster surveys
because its extent is poorly constrained by \rosat~data and its
flux is fainter than most survey flux limits.

\section{Optical Follow-up Observations}
\label{optical}

To reject the possibility of a relatively low-redshift cluster ($z \la
0.4-0.5$), we examined the location of \xmmu~in the second epoch
Digitized Sky Survey and found the region devoid of any galaxy
over-density.  To further constrain the redshift in an efficient
manner, we acquired relatively short-exposure images in the $R$
(1140s) and $z$ (480s) bands using VLT-FORS2 on 2 October 2003.
These images, combined with a subsequently obtained deep VLT-ISAAC
$K$$s$-band image (3600s, 9--11 December 2004), are shown as a
\mbox{2.5\arcmin\,$\times$\,2.5\arcmin}~ color composite in
\mbox{Figure \ref{fig:rzk}}.  The $Rz$ discovery imaging reveals a
significant over-density of faint, very red galaxies spatially
coincident with the peak of the extended X-ray emission.  Note that
the brightest cluster galaxy (BCG) has an extended surface brightness
profile typical of massive cluster cDs.

\begin{figure}
\epsscale{1.2}
\plotone{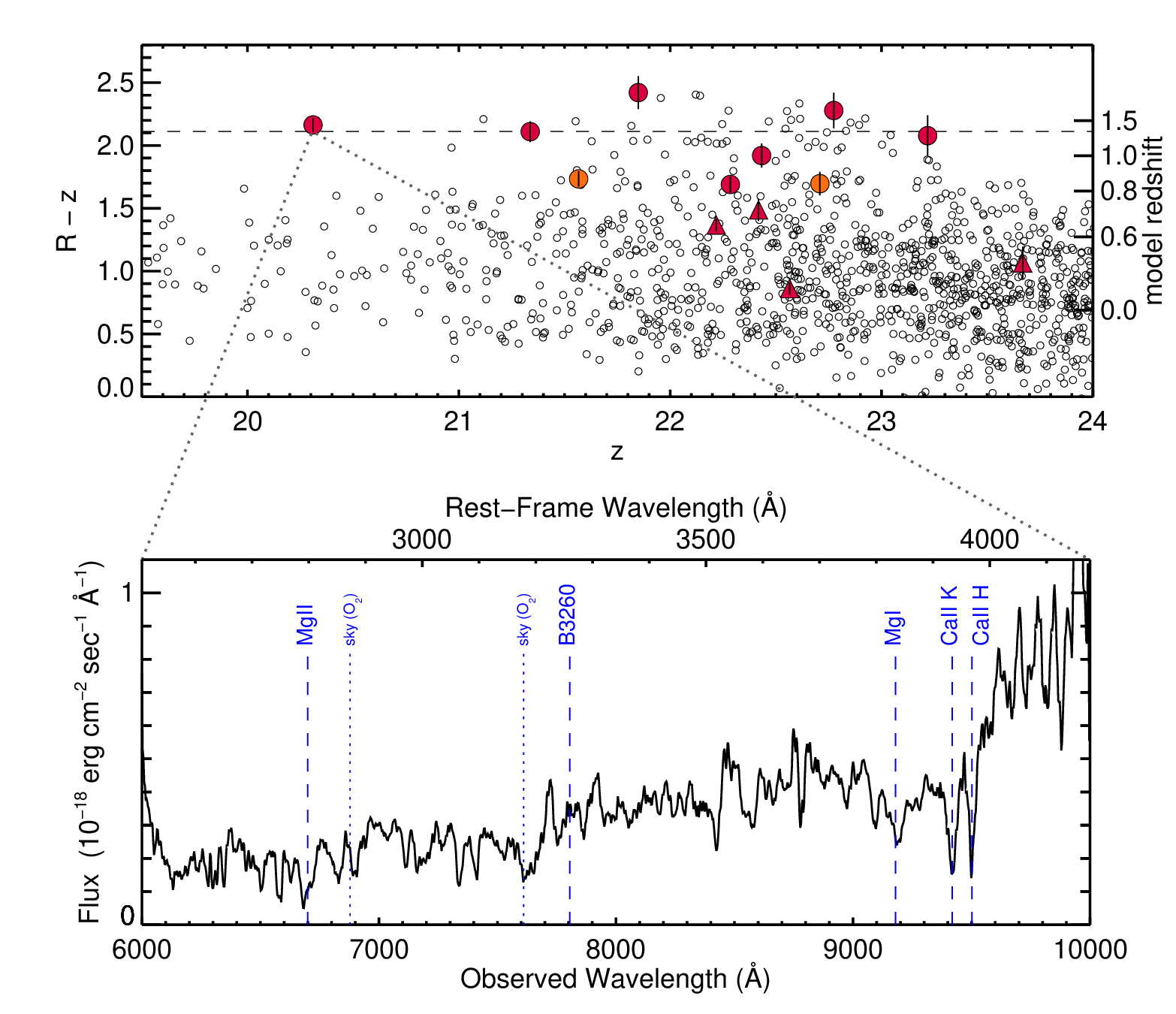}
\plotone{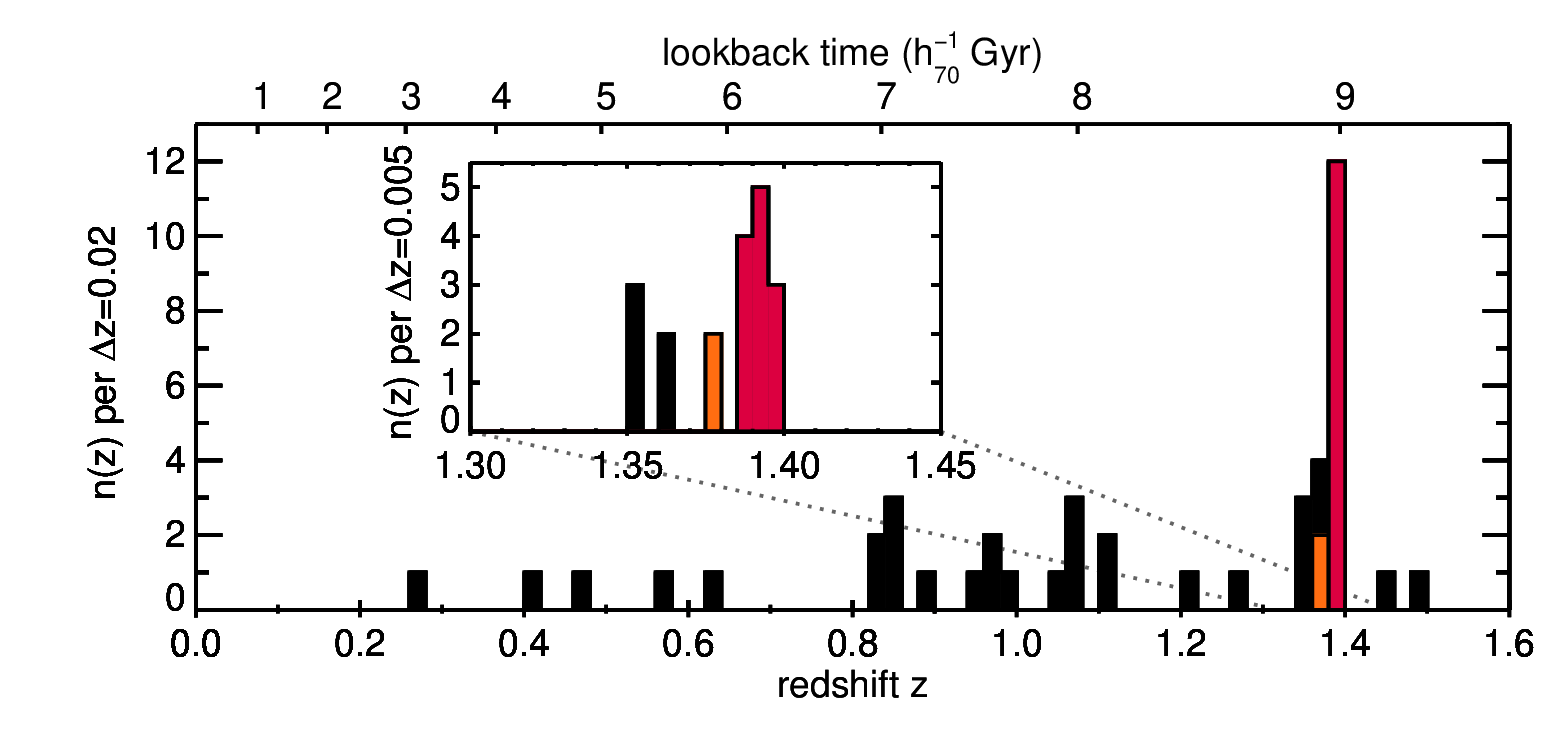}

\caption{{\em Top:} Color-magnitude diagram of the
7\arcmin\,$\times$\,7\arcmin~field around \xmmu.  Spectroscopically
confirmed cluster galaxies ($1.38 < z < 1.40$) are highlighted in red;
two galaxies at $1.37 < z < 1.38$ in orange (circle: absorption line
gal.; triangle: emission line gal.).  The horizontal line indicates
the predicted color of a $z=1.393$ cluster elliptical. {\em Middle:}
VLT-FORS2 spectrum of the brightest cluster galaxy ($z=1.3943 \pm
0.0003$, 4 hr integration). {\em Bottom:} Histogram of galaxy
redshifts measured in the VLT-FORS2 MXU observations of the 7\arcmin
$\times$ 7\arcmin~region around \xmmu~(same color scheme as above).
\label{fig:cmdspeczhist}}
\end{figure}

We plot in the top panel of \mbox{Figure \ref{fig:cmdspeczhist}} the
optical/NIR color-magnitude diagram of the galaxies detected in the
7\arcmin\,$\times$\,7\arcmin~$z$-band image.  The central cluster
galaxies clearly delineate the bright end of the cluster red sequence
at a color of $R-z \simeq 2.1$.  Given a realistic galaxy model, we
can use the location of the red sequence as a reliable distance
indicator \citep[e.g.,][]{Kodama1997,Gladders2000}.  Assuming cluster
ellipticals form via monolithic collapse at $z\approx 3$ and then
passively evolve to the observed redshift \citep*[e.g.,][]{Daddi2000},
we derive a color-redshift transformation indicated on the right-side
ordinate of the color-magnitude diagram.  Thus the observed red
sequence color of \xmmu~corresponds to a redshift of $z\sim 1.4$.

To confirm this very high redshift estimate, we obtained spectroscopic
data from two VLT-FORS2 MXU multi-object slit-masks observed on
\mbox{11 \& 15 October 2004}.  The result of a four-hour integration
on the BCG is shown in the middle panel of \mbox{Figure
\ref{fig:cmdspeczhist}}. We measure 12 secure redshifts in the range
$1.38 < z < 1.40$ with $<z>$=1.393 and a preliminary velocity
dispersion of $762 \pm 265$ \mbox{km s$^{-1}$}, corrected for
cosmological expansion (see histogram in bottom panel of \mbox{Figure
\ref{fig:cmdspeczhist}}).  These spectroscopically confirmed cluster
members are marked in red in \mbox{Figures \ref{fig:images} \&
\ref{fig:cmdspeczhist}}.

\section{Discussion}
\label{discussion}

Here we address a few fundamental characteristics of \xmmu~based on
the discovery datasets.  Due to the space limitations of this Letter,
we must defer a broad and in-depth discussion to a forthcoming paper.
At $z=1.393$ this cluster is the most distant bona-fide X-ray luminous
cluster known to date, and it lies well beyond the redshift range
($z=1.0-1.27$) of the only 5 previously known distant X-ray clusters
(all from {\em ROSAT}).  Note that by advancing to $z=1.4$, we can now
look 0.5 Gyr further back compared to the previous limit.  This is
quite significant given the relevant formation time scales ($1-3$
Gyrs) for the stellar populations in massive cluster galaxies.

With the cluster redshift in hand, we can derive additional X-ray
properties of \xmmu.  We estimate the temperature of the intra-cluster
medium via a joint fit to the M1+M2 spectra over the 0.3--4.5\,keV
energy range.  Assuming a thermal model (MEKAL) with a metallicity of
0.3 solar, a Galactic absorption column of $n_{\rm H}= 1.47 \times
10^{20}$ cm$^{-2}$ and fixing the cluster redshift at $z=1.393$, we
find $kT = 6.0^{+2.5}_{-1.8}$ keV.  As noted in \S\,\ref{x-ray}, the
cluster flux within a 50\arcsec~radius aperture based on this model is
\mbox{($2.6 \pm 0.2) \times 10^{-14}$ ergs cm$^{-2}$ s$^{-1}$}
\mbox{(0.5--2.0\,keV)}.  At the cluster redshift this aperture
corresponds to a physical radius of \mbox{421 $h_{70}^{-1}$ kpc}.  If
the cluster emission profile follows the typical King profile with a
slope of $\beta=0.7$ and a core radius of \mbox{140 $h_{70}^{-1}$
kpc}, then the photometry aperture encloses 74.9\% of the flux.  Thus
the total flux is \mbox{($3.6 \pm 0.3) \times 10^{-14}$ ergs cm$^{-2}$
s$^{-1}$} and the cluster rest-frame luminosity is \mbox{($3.0 \pm
0.2) \times 10^{44}$ $h_{70}^{-2}$ ergs s$^{-1}$}
\mbox{(0.5--2.0\,keV)}.  This high X-ray luminosity and the high rate
of spectroscopic identification (high richness) suggest that \xmmu~is
likely more massive than RDCS1252--29 (previously the most massive,
distant cluster known at $z=1.24$).

Examining the projected distribution of red galaxies, those with
colors similar to the spectroscopically confirmed cluster members, we
see in general that there is a higher density of objects to the north of
the cluster core versus to the south.  Note this asymmetry is
exaggerated in the confirmed members due to bias inherent to the design
of the spectroscopy slitmasks which were based on the relatively
shallow $Rz$ discovery images.  For example, an $\sim$30\arcsec~band
of right ascension beginning $\sim$15\arcsec~south of the X-ray peak
was undersampled due to the dithering pattern required to fill the gap
between the FORS2 CCDs.  The BCG and X-ray centroid are offset by
3.7\arcsec, or 31 $h_{70}^{-1}$ kpc at the cluster redshift, along a
northwest-southeast vector.  We must be cautious with our
interpretation until deeper X-ray data and additional spectroscopy are
available.  However, the alignment of the BCG--X-ray offset vector with
the northwest/southeast spurs in the X-ray morphology and the
filament of red galaxies leading out of the core to the northeast may
indicate a recent subcluster merger along this corridor.

\xmmu~is fairly isolated in redshift space (Figure
\ref{fig:cmdspeczhist}).  Note that there are two galaxies at
\mbox{$1.37 < z < 1.38$} shown in orange in \mbox{Figures
\ref{fig:images} \& \ref{fig:cmdspeczhist}}.  One of these ($z=1.379$)
is just outside the formal $3\sigma$ velocity boundary defining
cluster membership \mbox{($1.38 < z < 1.40$)}.  Both galaxies are
close in projection to the cluster core and likely part of the local
structure field immediately surrounding the main cluster.  The five
galaxies at \mbox{$1.35 < z < 1.37$} are roughly situated along a
declination band $\sim$2.5\arcmin~south of the cluster.  Four of these
fall within a 1.7\arcmin-diameter circle but there is no significant
X-ray emission in this region.

\section{Conclusions}
\label{conclusion}

\xmmu~(\mbox{$z=1.393$}) is the most distant X-ray--selected cluster
thus far discovered. Based on its high X-ray luminosity, ICM
temperature, and optical/NIR richness, this galaxy cluster is very
likely the most distant and most massive ($z > 1$) structure known to
date.  It provides an unprecedented opportunity to test models of the
evolution and formation of the most massive galaxies and clusters in
high-density environments at the largest look-back time currently
accessible.  Fundamental to this pursuit are high-quality datasets
including wide infrared coverage, high-resolution imaging from space,
optical spectroscopy \& dedicated X-ray follow-up.

A remarkable and exciting aspect of the discovery of \xmmu~is the
overall efficiency of telescope use from first detection to
spectroscopic confirmation.  Our experience demonstrates: 1) a
massive, $z=1.4$ cluster is easily detectable in a typical
\xmm~observation of $\sim$20\,ks, and 2) the red cluster sequence
provides a reliable distance indicator (out to at least $z=1.4$) which
can be measured in less than 30 minutes with a red-sensitive CCD on an
8m-class telescope.  In the search for $z>1$ clusters, the second
point is crucial for rejecting the large number of foreground clusters
and economizing the costly optical follow-up.  Given the relative ease
of discovery, we predict the detection of $z>1$ clusters will become
routine in the near future.

\acknowledgements

We thank the Paranal Science Operations \& ESO USG for efficiently
carrying out the VLT observations.  We are especially grateful to
Mario Nonino \& Chris Lidman for facilitating the rapid execution of
the DDT.  We thank Benoit Vandame \& Veronica Strazzullo
for supporting the Alambic/MVM image processing pipeline, Ricardo
Demarco for making his FORS2 MXU pipeline available to us, and
Emanuele Daddi for providing a stellar populations model. It is a
pleasure to thank Jimmy Irwin \& Gus Evrard for stimulating
discussions, and the referee, Stefano Borgani, for his thoughtful
review of this work.


\bibliographystyle{apj}
\bibliography{myrefs}

\end{document}